\documentclass[times,twocolumn,final]{elsarticle}

\usepackage{medima}
\usepackage{framed,multirow}

\usepackage{amssymb}
\usepackage{latexsym}
\usepackage{amsfonts}
\usepackage{amsmath}
\usepackage{mathtools}
\usepackage{url}
\usepackage{xcolor}

\usepackage{hyperref}
\usepackage{tabularx,booktabs}
\usepackage{multirow}
\usepackage{multicol}
\usepackage{xurl}
\usepackage{capt-of}
\usepackage{algorithm}
\usepackage{algpseudocode}
\usepackage{graphicx}
\usepackage{makecell}

\definecolor{newcolor}{rgb}{.8,.349,.1}

\newcommand{\E}{\mathbb{E}}
\newcommand{\R}{\mathbb{R}}

\newcommand{\bx}{\mathbf{x}}

\newcommand{\diffusionmodel}{{\boldsymbol{\epsilon}_{\theta}}}

\journal{Preprint}

\begin{document}

\verso{Antanas Kascenas \textit{et~al.}}

\begin{frontmatter}

\title{\Huge \centering \textbf{The role of noise in denoising models for anomaly detection in medical images}\tnoteref{tnote1}}%


\author[1,2]{Antanas \snm{Kascenas}\corref{cor1}}
\ead{antanas.kascenas@mre.medical.canon}
\author[3]{Pedro \snm{Sanchez}}
\author[1]{Patrick \snm{Schrempf}}
\author[1]{Chaoyang \snm{Wang}}
\author[1]{William \snm{Clackett}}
\author[1]{Shadia S. \snm{Mikhael}}
\author[1]{Jeremy P. \snm{Voisey}}
\author[1]{Keith \snm{Goatman}}
\author[1]{Alexander \snm{Weir}}
\author[2]{Nicolas \snm{Pugeault}}
\author[3,4]{Sotirios A. \snm{Tsaftaris}}
\author[1,3]{Alison Q. \snm{O'Neil}}

\address[1]{Canon Medical Research Europe, Bonnington Bond, 2 Anderson Pl, Edinburgh EH6 5NP, United Kingdom}
\address[2]{University of Glasgow, Glasgow G12 8QQ, United Kingdom}
\address[3]{University of Edinburgh, Kings Buildings, Edinburgh EH9 3FG, United Kingdom}
\address[4]{The Alan Turing Institute, London, United Kingdom}

\received{}
\finalform{}
\accepted{}
\availableonline{}
\communicated{}

\begin{abstract}
Pathological brain lesions exhibit diverse appearance in brain images, in terms of intensity, texture, shape, size, and location. Comprehensive sets of data and annotations are difficult to acquire. Therefore, unsupervised anomaly detection approaches have been proposed using only normal data for training, with the aim of detecting outlier anomalous voxels at test time. Denoising methods, for instance classical denoising autoencoders (DAEs) and more recently emerging diffusion models, are a promising approach, however naive application of pixelwise noise leads to poor anomaly detection performance. We show that optimization of the spatial resolution and magnitude of the noise improves the performance of different model training regimes, with similar noise parameter adjustments giving good performance for both DAEs and diffusion models.
Visual inspection of the reconstructions suggests that the training noise influences the trade-off between the extent of the detail that is reconstructed and the extent of erasure of anomalies, both of which contribute to better anomaly detection performance.
We validate our findings on two real-world datasets (tumor detection in brain MRI and hemorrhage/ischemia/tumor detection in brain CT), showing good detection on diverse anomaly appearances.  Overall, we find that a DAE trained with coarse noise is a fast and simple method that gives state-of-the-art accuracy. Diffusion models applied to anomaly detection are as yet in their infancy and provide a promising avenue for further research.

Code for our DAE model and coarse noise is provided at: \url{https://github.com/AntanasKascenas/DenoisingAE}.

\end{abstract}


\end{frontmatter}


\section{Introduction}

Anomaly detection is a fundamental task in medical image analysis, mimicking the initial review that a radiologist performs of imaging studies to identify abnormal regions which should be reviewed and characterized further. Supervised machine learning methods have shown promising results, however comprehensive supervised pathology detection methods require extensive and heterogeneous training sets which are costly to annotate and difficult to acquire. Conversely, unsupervised anomaly detection (UAD) methods require only identification of a healthy cohort of patients for training (therefore these methods are sometimes regarded as semi-supervised), after which they may be applied to detect out-of-distribution anomalous regions in test data.

Autoencoder deep learning methods have been commonly used for UAD in brain scans \citep{comparative}, relying on the assumption that normal data as seen during training will be reconstructed better than unseen anomalous -- potentially pathological -- regions. A classical approach is denoising autoencoders (DAEs) \citep{vincent2008extracting} in which corrupting noise is added to the input and the network must learn to remove the noise in order to reconstruct the original image. This training task of removing noise can be regarded as a proxy for the test time task of removing anomalies in order to reconstruct an image of normal appearance. It was shown in \cite{kascenas2021denoising} that for detection of brain tumors in MRI data, training with naive pixelwise noise gave poor anomaly detection performance, while training with coarse noise (see Algorithm \ref{alg:noise}) gave good performance. Following simple optimization of noise resolution and magnitude, a classical DAE outperformed more complex previous state-of-the-art models.

Our contributions are as follows:

\begin{enumerate}
    \item We take the simple and effective DAE that was proposed by  \cite{kascenas2021denoising} for brain anomaly detection in medical 2D MRI images, and investigate its application to 3D CT images with a range of anomalies, showing that optimal noise resolution and magnitude parameters are largely transferable between modalities and anomalies.
    \item We analyze noise type in the alternative denoising model paradigm of diffusion models, showing that similar adjustment of the type of noise gives accuracy gains also for this alternative denoising approach.
    \item We additionally analyze an alternative noise type (Simplex noise) which has been recently advocated by \cite{wyatt2022anoddpm}, showing our proposed coarse noise to be superior in most anomaly detection setups.
    \item Finally, since we consider training an anomaly detection algorithm in a practical setting where a large uncurated dataset of scans is available, we demonstrate that NLP analysis of radiology reports can be effectively used to select the training cohort of normal scans.
\end{enumerate}

\section{Related Work}

Anomaly detection is an open-ended task for which a variety of approaches have been proposed.

\subsection{Autoencoder approaches to anomaly detection}

Many modifications to the standard autoencoder pipeline have been proposed to improve performance on the task of anomaly detection, which has the potentially conflicting twin goals of reconstructing normal regions of the original brain scan with high fidelity, while reconstructing any anomalous regions with poor fidelity (in order to distinguish them).

Variational autoencoders (VAEs) \citep{vae1} constrain the latent bottleneck representation to follow a parameterized multivariate Gaussian distribution. \citet{vaegradient} further add a context-encoding task and combine reconstruction error with density-based scoring to obtain the anomaly scores, while \citet{restoration} use an iterative gradient descent restoration process at test time in restoration-VAE, replacing the reconstruction error with a restoration error to estimate anomaly scores.

Architectural changes have also been proposed. \citet{spatialae1,spatialae2} introduce convolutional autoencoders and higher capacity spatial bottlenecks instead of fully-connected (dense) bottlenecks to achieve better reconstruction. \citet{constrainedae} use constrained autoencoders to improve latent representation consistency in anomalous images at test time. Bayesian skip-autoencoders \citet{skipae} use skip connections with dropout to improve reconstruction and allow uncertainty to be measured via dropout stochasticity. \cite{baur2020scale} use scale-space autoencoders to compress and reconstruct different frequency bands of brain MRI using the Laplacian pyramid to achieve higher reconstruction fidelity. 

\begin{figure*}[t]
\centering
\includegraphics[width=1.0\textwidth]{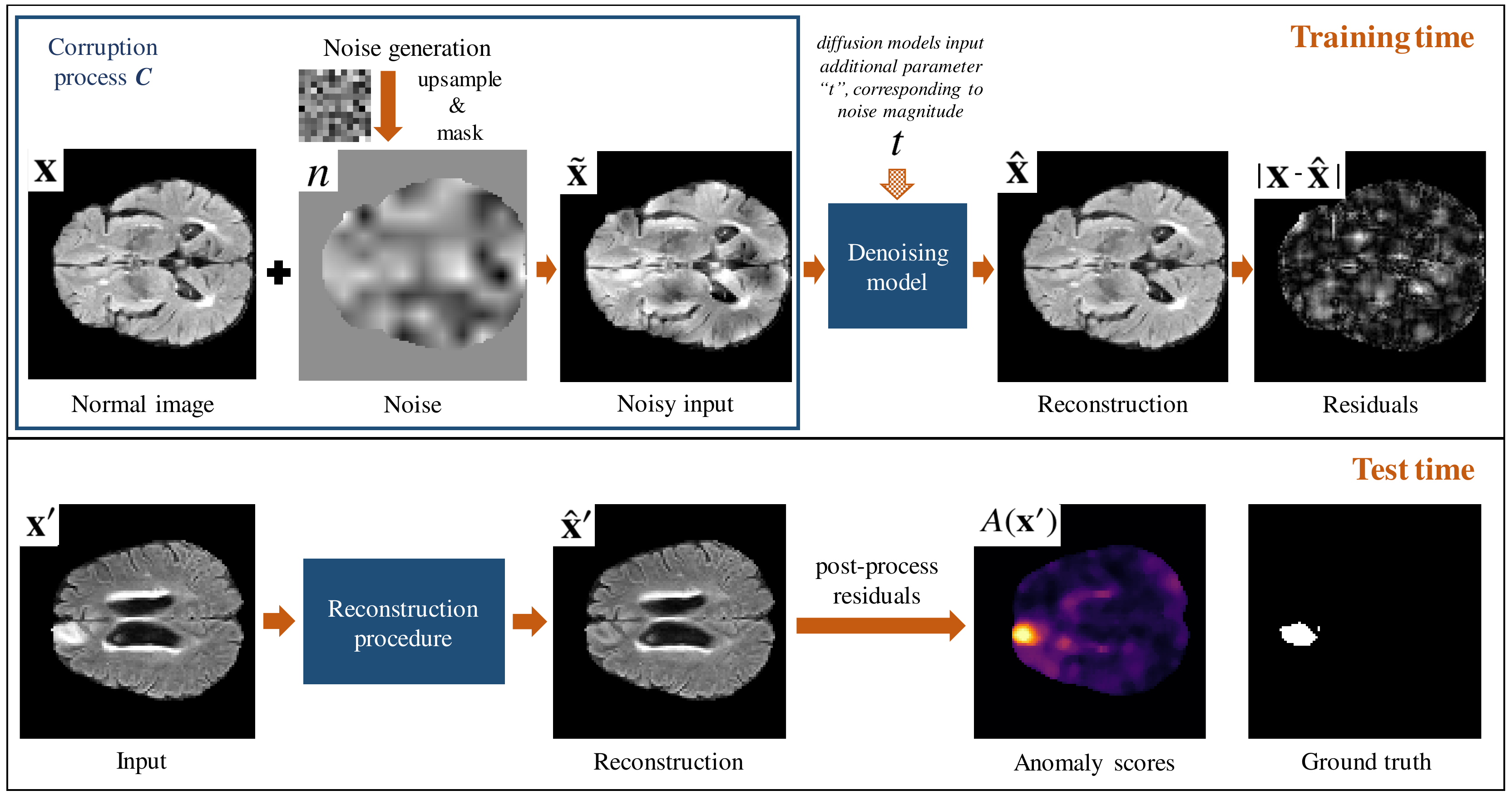}

\caption{Workflow for denoising anomaly detection methods.
During training (top), noise is added to the normal image, and the network is trained to reconstruct the original image.
At test time (bottom), different methods are applied to reconstruct and post-process the potentially anomalous input image to produce the anomaly score. For the simple denoising autoencoder (DAE) approach, the denoising model is applied once to the input and the anomaly score is simply the reconstruction residuals followed by median filtering. However, the diffusion models apply more complex iterative noise addition followed by iterative denoising to obtain the reconstruction.
}
\label{fig:pipeline}
\end{figure*}

The UAD autoencoder framework of encoder-decoder components and reconstruction error for anomaly scores has featured in more complex approaches. \citet{anogan} train a generative adversarial network called f-AnoGAN which reuses the generator and discriminator to train an autoencoder with both reconstruction and adversarial losses for the anomaly detection task. \citet{transformers} combine a vector quantized VAE (VQ-VAE) to encode an image with a transformer model to resample low-likelihood latent variables in order to produce reconstructions with fewer reproduced anomalies.

\citet{comparative} have performed an evaluation of some common autoencoder methods for anomaly detection in brain MRI, finding restoration-VAE \cite{restoration} and f-AnoGAN \cite{anogan} to be among the best. However, more recently \citet{thresholding} showed that most autoencoder-based MRI UAD methods can be outperformed by a simple thresholding baseline, applied to the FLAIR sequence after histogram equalization preprocessing. This training-free approach detected hyperintense brain tumor and multiple sclerosis lesions better than most UAD approaches that require healthy data to train.

Our work relies on the same principle of using reconstruction error for anomaly detection as most autoencoding methods but we use noise instead of architectural constraints to make the autoencoding training task non-trivial. 

\subsection{Denoising methods}
The above evaluations of medical anomaly detection methods largely omitted consideration of classical denoising autoencoders (DAEs) \citep{vincent2008extracting} and other methods exploiting noise, however a few approaches have shown promise. \cite{denoisingae} applied DAEs as pretraining for brain lesion detection with limited labels and for simple novelty detection using patch-based masking.
\cite{collin2021improved} use a DAE for anomaly detection in industrial vision with a stain noise model with randomized shape, color, size and location. \cite{bengs2021three} use 3D VAEs with spatial patches replaced with voxelwise noise to train an inpainting model for anomaly detection.
Generative diffusion models \citep{Ho2020DenoisingModels}, in which noise is added and removed over many iterations, have been used in the context of anomaly detection \citep{Pinaya2022,wyatt2022anoddpm} by assuming that models trained on only healthy data will fail during reconstruction of anomalous features.

Recently, \cite{kascenas2021denoising} showed that when noise coarseness and intensity are adjusted, a DAE can achieve competitive results for the detection of tumors in brain MRI images. Further, \citep{Giannis2022,wyatt2022anoddpm} have shown recently that diffusion models can be trained with degradation functions other than Gaussian noise. In fact, \cite{wyatt2022anoddpm} showed that using Simplex noise in diffusion models can significantly improve anomaly detection performance over traditional Gaussian noise.

In this paper, we examine this theme of the role of noise in anomaly detection, investigating and comparing types of noise and denoising methods in a common setting.

\section{Method}

In summary, we employ one of three types of noise (Gaussian, Simplex or coarse) to train neural network models to denoise healthy images. At test time, anomalies are detected via reconstruction error (see Figure \ref{fig:pipeline}). Below we describe this process in more detail.

\subsection{Denoising Models for Anomaly Detection}

Denoising neural networks $\diffusionmodel$ receive corrupted data $\tilde{\bx}$ as input and are trained to recover original (uncorrupted) data $\hat{\bx} = \diffusionmodel \left( \tilde{\bx} \right)$. We consider the corruption process to have a conditional distribution $C \left( \tilde{\bx} \mid \bx, n \right)$, degrading $\bx$ into $\tilde{\bx}$ with the injection of some noise $n$. Training a denoising neural network $\diffusionmodel$ with parameters $\theta$ can then be written as:
\begin{equation}
\theta^* = \underset{\theta}{\arg \min } ~ \E_{\bx \sim p_{\text{data}}, ~ \tilde{\bx} \sim C(\bx) } \left[ \left\| \diffusionmodel \left( \tilde{\bx} \right) - \bx \right\|^{2}\right].
\label{eq:training_denoising}
\end{equation}
The resulting network learns to reconstruct samples $\bx$ that belong to $p_{\text{data}}$.
However, we consider a distribution $p_{\text{anomaly}}$ which is similar to $p_{\text{data}}$ but contains features (\textit{anomalies}) that are not present in $p_{\text{data}}$. As shown by \citet{kascenas2021denoising}, an anomalous sample $\bx' \sim p_{\text{anomaly}}$ will not be reconstructed appropriately by the denoising network $\diffusionmodel$. The training and test pipelines are visualized in Figure \ref{fig:pipeline}.

The anomalies are detected by taking the absolute difference between the input data and the resulting reconstruction $|\bx' -  \hat{\bx}'|$.

\subsection{Denoising Autoencoder (DAE) approach}

We implement a simple denoising deep autoencoder neural network, and use reconstruction error to detect and localize anomalies at test time. The network has a U-Net \citep{unet} style architecture with skip connections which enables significantly better image reconstructions compared to bottleneck architectures such as the VAE (see \ref{apx:rec_comparison}). We note that any neural network architecture yielding dense predictions (e.g. segmentations) could be trained as a DAE.
Details of the network architecture and training procedure can be found in Section \ref{sec:implementation_details} and \ref{apx:architecture_details}. During training, we corrupt images according to:
\begin{equation}
  C\left( \tilde{\bx} \mid \bx \right) \implies \tilde{\bx} = \bx + \sigma  n,
\end{equation}
where $\sigma$ is the standard deviation which controls the intensity magnitude and $n$ is noise. Classically, $n$ is sampled from a Gaussian distribution $\mathcal{N}(0,\mathrm{I})$, but in Section \ref{sec:coarse_noise} we explore more efficient techniques in the context of anomaly detection.

At inference time, the DAE is used to localize anomalies by calculating pixelwise/voxelwise anomaly scores $A(x)$. If we denote the input image as $\bx$, the number of image channels as $M$ (e.g. for multiple imaging sequences or imaging modalities), a background mask of pixels with $\bx$ intensities across channels equal to 0 as $B$, the median filtering operation as $f$, and the reconstruction as $\hat{\bx}$, then the anomaly score can be defined as:
\begin{equation}    
 A(\bx) = f\left((1-B)\odot\sum^{M}_m\frac{|\bx_m - \hat{\bx}_m|}{M}\right) 
\end{equation}
No noise is used at test time.

\subsection{Diffusion model approach}

\begin{figure}[t]
\centering
\includegraphics[width=1.0\linewidth]{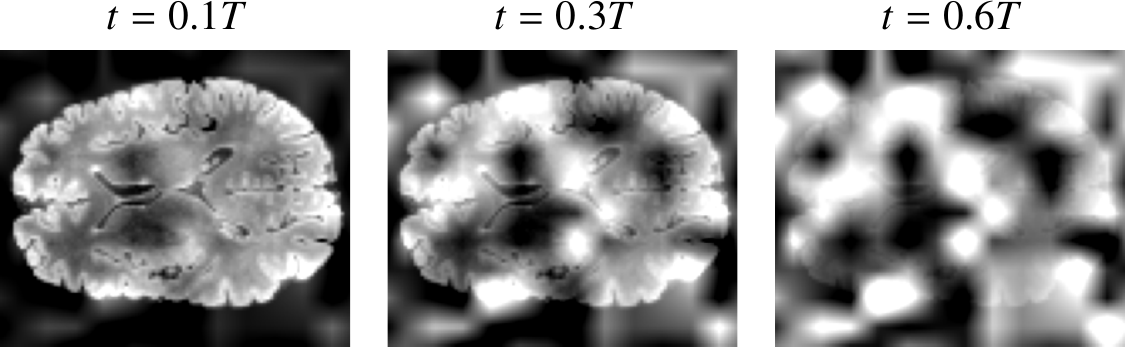}

\caption{
Diffusion model input at different timesteps $t$. Noise component is larger at further timesteps.
}
\label{fig:diffusion_noise}
\end{figure}

We next explore the diffusion model methods developed in \citet{Pinaya2022} and \citet{wyatt2022anoddpm}. Both methods follow a training strategy that was initially proposed by \citet{Ho2020DenoisingModels}. In contrast to the denoising model used in the DAE, diffusion denoising models are trained to predict the \emph{noise} rather than to reconstruct the original image itself. 
In particular, consider a model trained to find optimal parameters as 
\begin{equation}
\theta^* = \underset{\theta}{\arg \min } ~ \E_{\bx \sim p_{\text{data}} , ~ t \sim \mathcal{U}(0,T) , 
~ \tilde{\bx} \sim C(\bx, t)} \left[  \lambda(t) \left\| \diffusionmodel \left( \tilde{\bx} , t \right) - n \right\|^{2}\right],
\label{eq:trainingDDPM}
\end{equation}
where the timestep $t$ is sampled from a uniform distribution between $0$ and $T$ ($T$ is a hyperparameter which we set to 1000) and $\lambda(t)$ is a loss weighting term. Here the corruption process $C \left( \tilde{\bx} \mid \bx , t \right)$ depends also on $t$ which controls the strength of the corruption through $\alpha_t$ as described in \citet{song2021score}
, according to:
\begin{equation}
  C\left( \tilde{\bx} \mid \bx, t \right) \implies \tilde{\bx} = \sqrt{\alpha_t} \bx + \sqrt{1 - \alpha_t} n,
\end{equation}

\noindent{}where the coefficient $\alpha_t$ runs from $\alpha_0=1$ (original image) through to $\alpha_T=0$ (noise). Figure \ref{fig:diffusion_noise} shows examples of a corrupted image for different values of $t$. Training with multiple $t$ values corresponds to training with multiple noise magnitudes. Training with $C \left( \tilde{\bx} \mid \bx , t \right)$ has been extensively studied in the diffusion probabilistic modeling (DPM) literature \citep{Ho2020DenoisingModels}. For example, when using Gaussian noise, adding noise with high standard deviation causes the network to focus on coarse features while low standard deviation noise causes focus on texture and other high frequency detail. Most importantly, training to denoise at multiple magnitudes enables image \emph{generation}; we refer the reader to \citet{Ho2020DenoisingModels} for details on the image generation procedure. The ability of DPMs to denoise at different noise magnitudes as well as its generative power has inspired methods for anomaly detection using diffusion models \citep{Pinaya2022,wyatt2022anoddpm,sanchezDiffusionBrats}.


Once the diffusion denoising model has been trained, we investigate two inference techniques to detect anomalies:
\begin{enumerate}
    \item \textbf{Reconstruction} \citep{wyatt2022anoddpm} - In the AnoDDPM method, noise is injected at a selected magnitude; we use $t = 0.25T$ because this was found to be the best in \citet{wyatt2022anoddpm}. We then run the DDPM \citep{Ho2020DenoisingModels} iterative generation from $t = 0.25T \rightarrow 0$, using the noisy image as the starting point (i.e. 250 steps where T=1000). We follow \cite{wyatt2022anoddpm} in averaging the reconstructions across $5$ runs of this generation procedure.
    \item \textbf{KL divergence + inpainting} \citep{Pinaya2022} - In this method, noise is injected with different magnitudes (i.e. different $t \in \left[ 0.4T , 0.6T \right]$) and the difference is computed between the predicted output $\diffusionmodel \left( \tilde{\bx} , t \right)$ and the expected output $n$. A heatmap is obtained by averaging the difference images produced by different values of $t$; since DPMs have a probabilistic interpretation, \citeauthor{Pinaya2022} term this the Kullback–Leibler divergence. The KL divergence heatmap is binarized at a threshold corresponding to the 97.5 percentile value of the heatmap, to produce a mask for the region of interest. The masked region \emph{only} is then reconstructed by the model using the DPM i.e. ``inpainted'' \citep{lugmayr2022repaint} by running iterative generation from $t = 0.5T \rightarrow 0$.
    The final heatmap is the difference between the original and inpainted images.
\end{enumerate}

\begin{figure*}[t]
  \centering
  
  \includegraphics[width=1.0\textwidth]{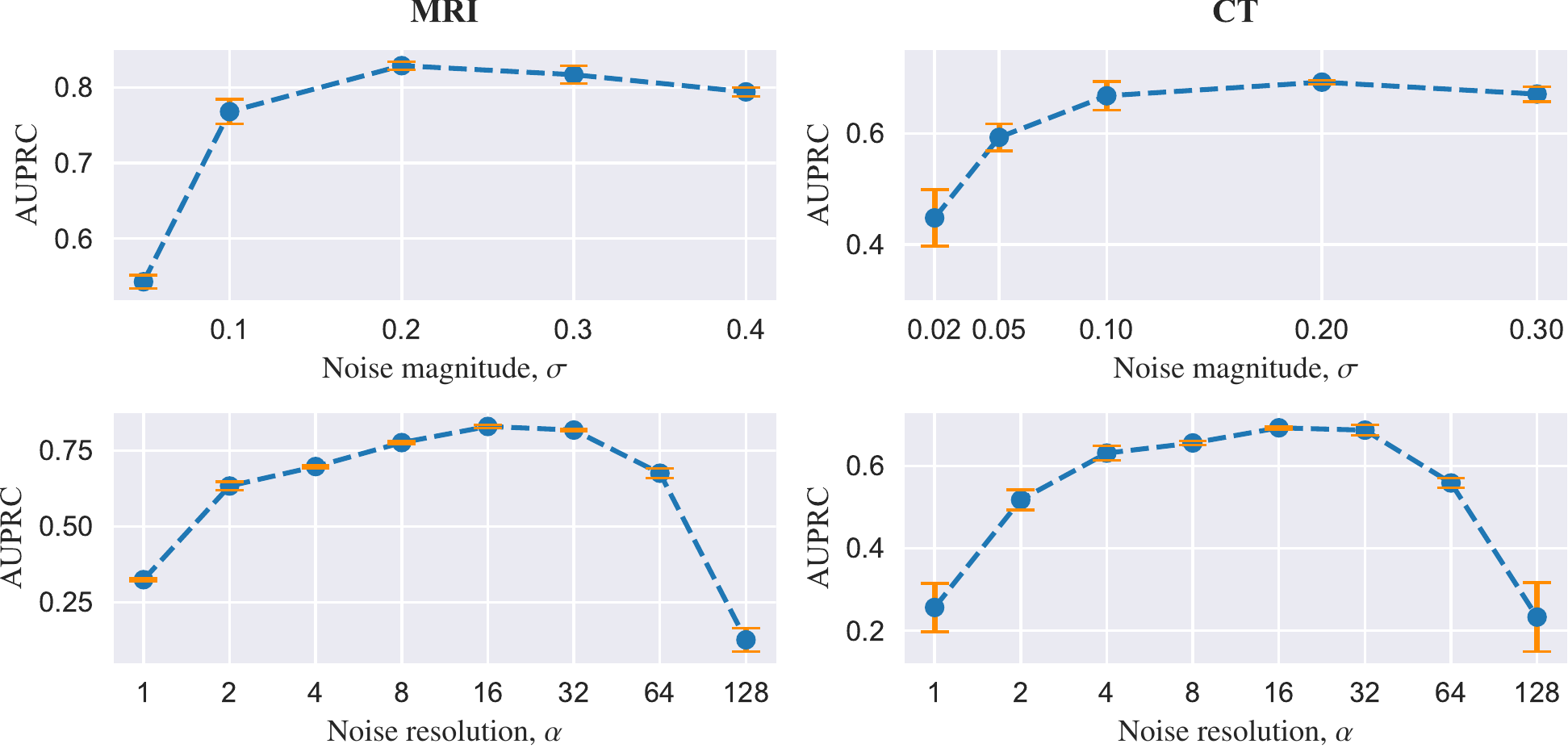}
  \caption{Noise coarseness and magnitude ablation results on BraTS head MRI (left) and iCAIRD head CT (right) data. Magnitude ablation uses noise resolution $\alpha=16$. Coarseness ablation uses $\sigma =0.2$. Error bars show standard deviation across three runs.}
  \label{fig:noise_ablation_full}
\end{figure*}

\begin{figure}[!htp]
  \centering
  \includegraphics[width=1.0\linewidth]{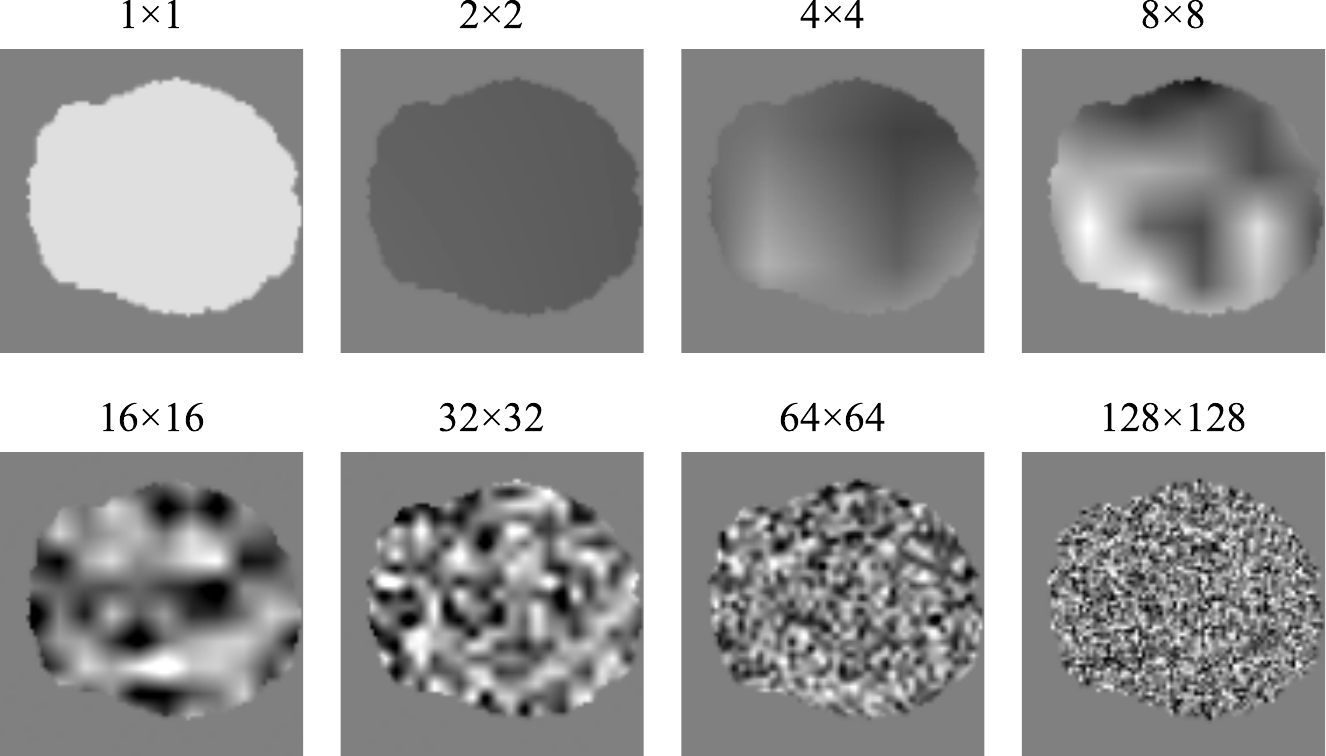}
  \captionof{figure}{Samples of 2D noise generated at different resolutions, from 1$\times$1 through to 128$\times$128. The background mask $B$ is also visible.}
  \label{fig:noise_samples}
\end{figure}

\subsection{Coarse noise generation}
\label{sec:coarse_noise}

It was shown in \cite{kascenas2021denoising} that training with lower resolution noise leads to better anomaly detection than naive pixelwise Gaussian noise for a DAE detecting brain tumors in brain MRI data; in this paper we are interested to further investigate the impact of the type of noise with different data and denoising models. We generate the lower resolution (``coarse'') noise by sampling random pixelwise Gaussian noise at a low resolution and bilinearly (trilinearly for 3D) upsampling it to the input resolution. We then randomly translate the generated noise to avoid consistent upsampling patterns. 
See Figure \ref{fig:noise_samples} for examples of generated noise and Algorithm \ref{alg:noise} for the pseudocode.
\begin{algorithm}[t]
\caption{\\Generation of noise with spatial resolution $\alpha$, and output shape $a\times b\times c$}\label{euclid}
\begin{algorithmic}[1]
\Procedure{Noise}{$\alpha, a, b, c$}
\State $n\sim N(0, \mathrm{I}) \in \R^{\alpha, \alpha, \alpha} $
\State $n\gets \text{upsample}(n, (a, b, c))$\Comment{Bilinearly upsample}

\State $x\sim \mathcal{U}(0, a)$\Comment{Uniformly sample in range $(0, a)$}
\State $y\sim \mathcal{U}(0, b)$
\State $z\sim \mathcal{U}0, c)$
\State $n\gets \text{translate}(n, (x, y, z))$ \Comment{Randomly translate}
\State \textbf{return} $n$\Comment{The generated noise is n}
\EndProcedure
\end{algorithmic}
\label{alg:noise}
\end{algorithm}

To examine the effect of noise, we take the DAE and vary the noise resolution $\alpha$ and the standard deviation $\sigma$ used for generating Gaussian noise before upsampling (see Figure \ref{fig:noise_ablation_full}) on the two datasets described later in Section \ref{sec:datasets}. We find that a reasonably coarse noise is critical, as DAE models trained using standard pixel-level noise (e.g. generated at $128\times128$ resolution for $128\times128$ pixel 2D head MRI slices) or using the opposite extreme of image-level noise (i.e. generated at $1\times1$ resolution) perform significantly worse. DAEs appear to be less sensitive to the magnitude of the noise ($\sigma$ of the generating Gaussian distribution). The noise parameters are robust and transfer well between modalities (i.e. from MRI to CT) and pathologies (e.g. tumor to hemorrhage) and 2D to 3D as long as the image field-of-view and resolution are comparable i.e. we used 1.62mm$^2$/pixel for 2D MRI and 2mm$^3$/voxel for 3D CT.

For diffusion models, we similarly modify the corruption process $C$, adopting the noise generation process described in Algorithm \ref{alg:noise} instead of pixel-wise Gaussian noise both during training and when applying each of the inference techniques.

\section{Datasets}
\label{sec:datasets}

We evaluate anomaly detection in 2D brain MRI slices and 3D brain CT volumes using the two datasets described below.

\subsection{BraTS challenge dataset: Brain MRI}
We evaluate the anomaly detection performance on the surrogate task of brain tumor segmentation using data from the BraTS 2021 challenge \citep{brats1, brats2, brats3}. This data comprises native (T1), post-contrast T1-weighted (T1Gd), T2-weighted (T2), and T2 Fluid Attenuated Inversion Recovery (FLAIR) modality volumes for each patient from a variety of institutions and scanners.

\subsubsection{Selecting the training and test data}

We randomly split the dataset into 938 training, 62 validation, and 251 test patients. During training, we use only slices that do not contain any tumor pixels, under the assumption that these non-tumor slices represent healthy tissue. At test time, we consider the union of the tumor sub-region labels to be the anomalous regions.

\subsubsection{Preprocessing}

The data has already been co-registered, skull-stripped and interpolated to the same resolution. Labels are provided for tumor sub-regions: the GD-enhancing tumor, the peritumoral edema, and the necrotic and non-enhancing tumor.

 For the data input to the models, we stack all four modalities at the channel dimension for each patient. We normalize (rescale) the pixel intensity values in each modality of each scan by dividing by the 99th percentile foreground voxel intensity. Values are scaled to a range of $[-1,1]$ for diffusion methods and $[0,1]$ otherwise. All slices are downsampled to a resolution of 128$\times$128 (1.62mm/pixel).

\subsection{iCAIRD GG\&C NHS dataset: Head CT}

We use head CT scans obtained through a collaboration with the Industrial Centre for Artificial Intelligence Research in Digital Diagnostics (iCAIRD)\footnote{\url{https://icaird.com}}. The data has been sourced from hospitals in the Greater Glasgow \& Clyde (GG\&C) area in Scotland and comprises all patients who were diagnosed with a stroke in the period 2013-2018. The data is pseudonymised and we obtain access onsite via the West of Scotland Safe Haven within NHS Greater Glasgow and Clyde via the Safe Haven Artificial Intelligence Platform (SHAIP) \citep{shaip}. We have obtained ethical approval to use this data\footnote{West of Scotland Safe Haven ethical approval number GSH19NE004}.

The data was originally collected by identifying hospital admissions which were assigned International Classification of Diseases (ICD-10)\footnote{\url{https://www.who.int/standards/classifications/classification-of-diseases}} codes relating to stroke diagnoses, and then selecting medical data from the stroke event hospital admission as well as the documentation from 18 months prior and all prior images held at the GG\&C. In total, the dataset contains information about 15,882 stroke events from 10,143 patients and includes CT images, radiology reports, clinical documents and structured clinical data. We use 16,559 head CT images available from 7,122 patients for the purpose of this work and refer to this as the iCAIRD dataset.

\subsubsection{Radiology report NLP for normal scan selection}

\begin{table}[t]
  \caption{Data filtering steps towards obtaining a healthy training set for anomaly detection.}
  \label{tab:normals_filtering}
  \centering

  \begin{tabularx}{\linewidth}{Xrr}
  \toprule

  \bfseries Filtering step & \bfseries Images & \bfseries Patients \\
  
  \cmidrule(r){1-1} \cmidrule(lr){2-2} \cmidrule(lr){3-3} 

  Initial Data cohort & 16,559 & 7,122 \\
  After filtering on report labels from \cite{patrick_templates} & 2,350 & 1,788 \\ 
  After filtering out follow-up scans & 1,020 & 961 \\
  After rapid manual image review & 996 & 939 \\
  \bfseries Healthy training set & \bfseries 804 & \bfseries 757 \\
  
  \bottomrule
  \end{tabularx}
\end{table}

Identification of normal scans by manual examination of this large dataset would be time-consuming. Fortunately, corresponding free text radiology reports are available for most of the head CT images in the iCAIRD dataset. The reports vary in depth and exposition reflecting the style and seniority of the reporting radiologists, but generally describe the radiographic findings and clinical impressions in the associated CT images. We use this information to identify and exclude abnormal scans from our training set. However, comprehensive manual examination of radiology reports, while faster than examination of images, would still be slow. Therefore, we leverage a previously developed automatic deep learning model \citep{patrick_attention,patrick_templates} which was trained on 357 manually labeled non-contrast head CT radiology reports and outputs labels for 14 radiographic findings and 19 clinical impressions (see \ref{apx:patrick_labels} for the list of labels). Each label is assigned one of the 4 classes: \emph{positive}, \emph{negative}, \emph{uncertain} or \emph{not mentioned}.

\subsubsection{Selecting the training data}

\vspace{5pt}\noindent{}\textbf{Defining Normal vs Abnormal:} We aim to obtain a training set that is as healthy as possible in order to detect as many anomalies as possible at test time. However, since the dataset is from an elderly stroke population (mean age of 72 years), reports without any positive findings (labels) are rare. Therefore, there is a trade-off between how aggressively we filter versus the size of the final training set.
Hence, we include scans for which the associated reports contain only findings/impressions that are commonly found in an elderly population, specifically calcification, atrophy, cerebral small vessel disease and hypodensity (the latter is most commonly associated with atrophy and small vessel disease). Applying this more generous definition of ``Normal'' leaves a set of 2350 scans from 1788 patients (see Table \ref{tab:normals_filtering}).

\vspace{5pt}\noindent{}\textbf{Filtering out follow-up scans:} Upon closer manual inspection we find that many reports are non-exhaustive (note these are free text rather than structured reports), appearing not to list all of the findings present in the scan. This most commonly occurs for follow-up scans where the associated report assumes knowledge of earlier scan reports, usually not explicitly re-listing all findings. An example such report would be ``\emph{No progression compared to previous scan from 10/22/2021.}''. Thus, absence of positive or uncertain labels does not necessarily equate to absence of pathology. Therefore we further filter down the remaining cases using keywords and pattern matching using spaCy \citep{spacy}, removing reports which contain references to previous imaging and comparisons. This keyword filtering leaves 1020 scans from 961 patients.

\vspace{5pt}\noindent{}\textbf{Rapid manual image review for obvious anomalies:} Finally, we perform a rapid manual review by non-experts which eliminates a further 24 scans mostly containing processing issues (e.g. bone reconstruction, significant artifact, significantly degraded scan quality). We use 804 scans from the remaining 996 cases as our healthy training data. 

\subsubsection{Selecting and annotating the test data}
\label{sec:pathology-gt}

In addition to the filtered healthy training data, we selected and annotated a separate set of scans with hemorrhages, ischemia and tumors to quantitatively evaluate the methods. The annotation workflow consisted of several steps: curation, annotation, review and quality assurance. Further details are provided in \ref{app:icaird-test-set}. The resulting data was split into Test and Training sets as described below.

\vspace{5pt}\noindent{}\textbf{Test set:} The test set contains voxelwise annotations for 114 scans of which 104, 23 and 4 contain hemorrhage, ischemia and tumor ground truth respectively. We use the union of the three pathologies for evaluating the anomaly detection methods.

\vspace{5pt}\noindent{}\textbf{Training data for supervised baselines:} We further reserve 129 scans annotated with 116 hemorrhage, 30 ischemia and 6 tumor annotations for training the supervised baselines.

\subsubsection{Preprocessing}

We rigidly register the CT scans to a reference volume and crop to a fixed field-of-view which includes only the head region of the scan. Volumes are then resampled to 2mm$^3$ resolution and windowed to Hounsfield Unit (HU) values from 0 to 80. As for the MRI data, intensities are rescaled to a range of $[-1,1]$ for diffusion methods and $[0,1]$ otherwise. We use random flipping and affine transformation data augmentation for training of all methods.

\subsection{qure.ai CQ500: Head CT}

We use the CQ500 dataset from qure.ai \citep{qureai} for qualitative evaluation (see Figure \ref{fig:qualitative_ct}) of the head CT methods as the data contains similar pathologies (i.e. hemorrhages, ischemia). This dataset does not, however, contain any voxel-level ground truth and could not be used for quantitative evaluation.

\section{Baselines}
\label{sec:setup}

We compare against a range of common reconstruction-error based methods as well as providing supervised segmentation model results trained using ground truth for context. 

\subsection{2D brain MRI baselines}

We compare the denoising anomaly detection model performance against four methods. Firstly, we implement a standard VAE \citep{vae1} and f-AnoGAN \citep{anogan} models with pixelwise reconstruction error as the anomaly scores. Secondly, we use the same VAE model but implement an iterative gradient-based restoration process \citep{restoration} to produce restoration images.
Finally, we apply the simple thresholding approach from \citet{thresholding} modified to use median filtering as proposed by \citep{kascenas2021denoising}. We use the hyperparameters from the original works for the deep learning methods but tune manually where necessary to improve training stability and anomaly detection performance.

\subsection{3D brain CT baselines}

We compare the denoising anomaly detection model performance on 3D head CT data against two reconstruction-error based methods: VAE reconstruction \citep{vae1} and VAE restoration \citep{restoration}.

\section{Implementation details}
\label{sec:implementation_details}

\subsection{Noise}

Coarse noise is generated by sampling random Gaussian pixelwise noise at resolutions of $16\times16$ and $16\times16\times16$ for 2D and 3D respectively, before bilinearly/trilinearly upsampling to the input resolution of $128\times128$ for 2D brain MRI and $80\times112\times88$ for 3D head CT. The generated noise is then randomly translated to randomize the centers of the coarse noise peaks that may occur due to upsampling from very low resolutions. Noise is generated independently for each image modality in the case of 2D MRI. We investigate the parameters of the noise, as reported in Section \ref{sec:coarse_noise} (see Figure \ref{fig:noise_ablation_full}).

Simplex noise is generated using the implementation provided by \cite{wyatt2022anoddpm}\footnote{\url{https://github.com/Julian-Wyatt/AnoDDPM}}. For DAE experiments with Simplex noise, we scale the generated noise magnitude by a factor of $0.2$.

\subsection{Denoising autoencoder}
For 2D MRI data, we use a U-Net \citep{unet} encoder-decoder architecture with three downsampling/upsampling stages. Each encoder stage consists of two weight-standardized convolutions \citep{ws} with kernel sizes of 3 and 64, 128, 256 output channels for the three stages respectively followed by Swish activations \citep{swish} and group normalization \citep{groupnorm}. Average $2\times2$ pooling is used for downsampling. The decoder architecture mirrors the encoder in reverse, using transposed convolutional layers for upsampling. Architecture visualization and further details can be found in \ref{apx:architecture_details}.

For 3D head CT data, we use an analogous architecture in 3D with three downsampling/upsampling stages and 32, 64, 128 output channels for the three stages respectively.

We use mean $L^2$ reconstruction loss in the foreground as the training objective.
2D DAE Models are trained for 67,200 iterations with a batch size of 16 slices using Adam \citep{adam} with a cosine annealed \citep{cosine_anneal} maximum learning rate of 0.0001 with a period of 200 iterations. 3D DAE Models are trained for 25,600 iterations with a batch size of 3 volumes using Adam with OneCycleLR learning rate schedule \citep{onecyclelr} with a maximum learning rate of 0.001.

\subsection{Diffusion model}

We implement a diffusion model with a U-Net-like architecture based on implementation provided by \cite{dhariwal2021diffusion} which includes residual layers, global attention, dropout and a projection of the timestep embedding to each residual block. We use $T=1000$ diffusion steps with linear noise schedule training models to predict the noise and optimizing the mean squared loss between the noise which was used for sampling and the predicted noise.

For 2D MRI data we use the AdamW optimizer \citep{adamw} with a learning rate of 0.0001 and weight decay of 0.01 with a batch size of 64. Model weights are averaged by taking the exponential moving average (EMA) with a rate of 0.9999. The 2D U-Net architecture and diffusion training code can be found on github \footnote{\url{https://github.com/vios-s/Diff-SCM}}.

For 3D CT data we use the AdamW optimizer with a OneCycleLR learning rate schedule \citep{onecyclelr} with a maximum learning rate of 0.0001 and batch size of 4. Model weights are averaged by taking the EMA with a rate of 0.95.

\subsection{VAE reconstruction}
VAE models use a similar architecture to their DAE counterparts. Skip connections are removed and a bottleneck with dimensionality of 128 is added. For the training objective, we compute the sum of mean $L^2$ reconstruction error and KL-divergence with a weight of $\beta=0.001$. We use the same training procedure and anomaly score formula as for their DAE counterparts.

\begin{figure}[t]
  \centering
  \includegraphics[width=1.0\linewidth]{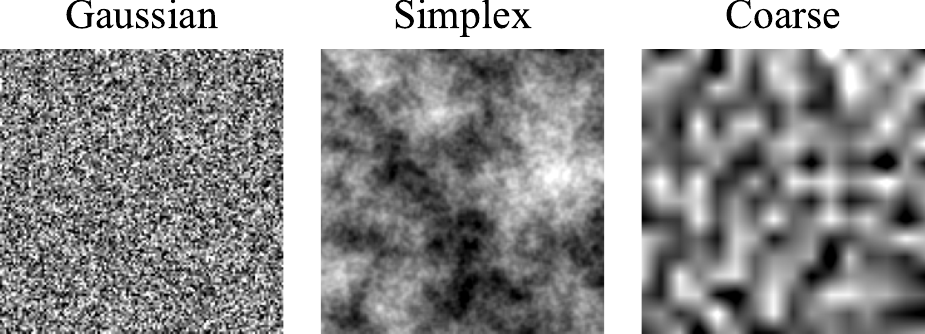}
  \caption{The three noise types tested to train diffusion models.}
  \label{fig:noise_type}
\end{figure}

\begin{table}[t]
  \centering

  \caption{Relationship between DAE and different diffusion anomaly detection inference methods and noise used for model training. For the method of \citep{Pinaya2022} we include also the results of the intermediate KL step. Numbers show area under the precision-recall curve (AURPC). Mean results reported across 3 runs $\pm$ standard deviation.}
  \begin{tabularx}{\linewidth}{Xccc}
  \toprule
   & \multicolumn{3}{c}{\bfseries Noise} \\
  \cmidrule(lr){2-4}
  \bfseries Inference method & \bfseries Gaussian & \bfseries Simplex & \bfseries Coarse\\
  \midrule
  \emph{2D Head MRI} &  &  & \\
  \makecell*[l]{Reconstruction\\ \citep{wyatt2022anoddpm}} & \makecell{0.197 \\ {\scriptsize $\pm$ 0.032}} & \makecell{0.464 \\ {\scriptsize $\pm$ 0.048}} & \bfseries \makecell{0.653 \\ {\scriptsize $\pm$ 0.063}}\\
  \makecell*[l]{KL + inpainting \\ \citep{Pinaya2022}} & \makecell{0.305 \\ {\scriptsize $\pm$ 0.008}} & \makecell{0.640 \\ {\scriptsize $\pm$ 0.020}}  & \bfseries \makecell{0.689 \\ {\scriptsize $\pm$ 0.028}}\\
  \makecell*[l]{$\rightarrow$ KL step only \\ \citep{Pinaya2022}} & \makecell{0.258 \\ {\scriptsize $\pm$ 0.009}} & \makecell{0.675 \\ {\scriptsize $\pm$ 0.035}}  & \bfseries \makecell{0.796 \\ {\scriptsize $\pm$ 0.022}} \\
  \makecell*[l]{ DAE \\ \citep{kascenas2021denoising}} & \makecell{0.325 \\ {\scriptsize $\pm$0.004}} & \makecell{0.723 \\ {\scriptsize $\pm$0.019}}  & \bfseries \makecell{0.833 \\ {\scriptsize $\pm$0.005}} \\
  \midrule
  \emph{3D Head CT} &  &  & \\
  \makecell*[l]{Reconstruction\\ \citep{wyatt2022anoddpm}} & \makecell{0.312 \\ {\scriptsize $\pm$0.027}} & \bfseries \makecell{0.623 \\ {\scriptsize $\pm$0.004}} & \makecell{0.573 \\ {\scriptsize $\pm$0.012}}\\
  \makecell*[l]{KL + inpainting \\ \citep{Pinaya2022}} & \makecell{0.069 \\ {\scriptsize $\pm$0.003}} & \makecell{0.357 \\ {\scriptsize $\pm$0.005}}  & \bfseries \makecell{0.512 \\ {\scriptsize $\pm$0.005}} \\
  \makecell*[l]{$\rightarrow$ KL step only \\ \citep{Pinaya2022}} & \makecell{0.098 \\ {\scriptsize $\pm$0.005}} & \makecell{0.432 \\ {\scriptsize $\pm$0.005}}  & \bfseries \makecell{0.629 \\ {\scriptsize $\pm$0.002}} \\
  \makecell*[l]{ DAE \\ \citep{kascenas2021denoising}} & \makecell{0.233 \\ {\scriptsize $\pm$0.084}} & \makecell{0.611 \\ {\scriptsize $\pm$0.038}}  & \bfseries \makecell{0.693 \\ {\scriptsize $\pm$0.004}} \\
  \bottomrule
  \end{tabularx}
  \label{tab:results_diffusion_both}
\end{table}

\subsection{VAE restoration}
Using the VAE model described above, we implement a restoration method \citep{restoration} to produce the anomaly scores. We perform the restoration procedure using 100 iterations on individual slices/volumes basing our implementation on public source code\footnote{\url{https://github.com/yousuhang/Unsupervised-Lesion-Detection-via-Image-Restoration-with-a-Normative-Prior}}. Note that due to the iterative nature of the restoration procedure it takes significantly longer (approx. $\times$100) to produce predictions compared to the single inference iteration DAE/VAE reconstruction.

\subsection{f-AnoGAN}

We adapt the original public implementation\footnote{\url{https://github.com/tSchlegl/f-AnoGAN}} for the brain MR data task as follows. We use an additional generator, discriminator and encoder block to account for the higher resolution. Strided convolutions and transposed convolutions are used for downsampling and upsampling respectively. We use a batch size of 32 and learning rates of 0.001, 0.001, 0.00001 for the generator, discriminator and encoder respectively. The encoder was trained using $\kappa=1\times 10^{-8}$.

\subsection{Thresholding} 

We follow \cite{thresholding} to obtain results for the thresholding baseline but omit the connected component filtering as we have found median filtering to be more effective and computationally efficient \citep{kascenas2021denoising}. FLAIR sequence volumes are used as the anomaly score volumes, following processing by histogram equalization of the foreground (i.e. excluding surrounding air) and median filtering.

\subsection{Supervised segmentation baselines}

We train supervised baselines to provide context on the expected performance range. The supervised head MRI baseline was trained using a 2D U-Net model with the same architecture as the DAE using 938 annotated volumes with tumor ground truth. The supervised head CT baseline was trained using the nnU-Net package \citep{nnunet} using 129 annotated volumes with hemorrhage, ischemia and tumor ground truth as described in Section \ref{sec:pathology-gt}.

\subsection{Postprocessing}

We use the same postprocessing in all tested methods. We apply median filtering with a kernel size of 5 which effectively reduces the false positives in the anomaly score heatmaps as shown in \cite{kascenas2021denoising} by filtering out insignificant reconstruction noise.

\section{Results}
\label{sec:results}

We now examine the difference in performance between noise types (Gaussian, Simplex, Coarse), between models (VAE, DAE, Diffusion models), across different modalities (MRI and CT), and between noise resolutions applied to different anomaly sizes. We finally inspect the model outputs to observe the difference in behavior qualitatively.

\begin{table}[b]
  \centering
  
  \caption{Pathology detection performance as evaluated on BraTS Head MRI Tumor test set. Metrics are the test set wide pixel-level area under the precision-recall curve (AUPRC) and ideal Dice score ($\lceil$Dice$\rceil$). Mean results reported across 3 runs $\pm$ standard deviation.}
  \begin{tabularx}{\linewidth}{Xccc}
  \toprule
  \bfseries Method & \bfseries AUPRC & \bfseries $\lceil$Dice$\rceil$ \\
  \midrule
  Thresholding  & 0.798 & 0.749 \\
  \midrule
  f-AnoGAN & 0.365{\scriptsize $\pm$0.024} & 0.449{\scriptsize $\pm$0.014} \\
  VAE (reconstruction) & 0.555{\scriptsize $\pm$0.004} & 0.548{\scriptsize $\pm$0.003} \\
  VAE (restoration) & 0.750{\scriptsize $\pm$0.006} & 0.689{\scriptsize $\pm$0.005} \\
  {Diffusion (reconstruction)} & 0.653{\scriptsize $\pm$0.063} & 0.610{\scriptsize $\pm$0.060}\\
  {Diffusion (KL + inpainting)} & 0.689{\scriptsize $\pm$0.028} & 0.675{\scriptsize $\pm$0.015} \\
  {$\rightarrow$ KL step only} & 0.796{\scriptsize $\pm$0.022} & 0.723{\scriptsize $\pm$0.013} \\
  DAE ($\alpha=16, \sigma=0.2$)  & \textbf{0.833{\scriptsize $\pm$0.005}} & \textbf{0.773{\scriptsize $\pm$0.004}} \\
  \midrule
  U-Net (supervised) & 0.972{\scriptsize $\pm$0.001} & 0.914{\scriptsize $\pm$0.002} \\

  \bottomrule
  \end{tabularx} 
  \label{tab:results_mri}
\end{table}

\begin{table}[b]
  \centering
  \caption{Pathology detection performance as evaluated on iCAIRD Head CT Hemorrhage/Ischaemia/Tumour test set. Metrics are the test set wide pixel-level area under the precision-recall curve (AUPRC) and ideal Dice score ($\lceil$Dice$\rceil$).  Mean results reported across 3 runs $\pm$ standard deviation.}
  \begin{tabularx}{\linewidth}{Xccc}
  \toprule
  \bfseries Method & \bfseries AUPRC & \bfseries $\lceil$Dice$\rceil$ \\
  \midrule
  VAE (reconstruction) & 0.382{\scriptsize $\pm$0.003} & 0.432{\scriptsize $\pm$0.005} \\
  VAE (restoration)& 0.542{\scriptsize $\pm$0.012} & 0.537{\scriptsize $\pm$0.011} \\
  Diffusion (reconstruction) & 0.573{\scriptsize $\pm$0.012} & 0.600{\scriptsize $\pm$0.013}\\
  Diffusion (KL + inpainting) & 0.512{\scriptsize $\pm$0.005} & 0.547{\scriptsize $\pm$0.008} \\
  $\rightarrow$ KL step only & 0.629{\scriptsize $\pm$0.002} & 0.608{\scriptsize $\pm$0.003} \\ 
  DAE ($\alpha=16, \sigma=0.2$) & \textbf{0.693{\scriptsize $\pm$0.004}} & \textbf{0.674{\scriptsize $\pm$0.003}} \\

  \midrule
  nnU-Net (supervised) & 0.817{\scriptsize $\pm$0.002} & 0.786{\scriptsize $\pm$0.004} \\
  \bottomrule
  \end{tabularx}
  \label{tab:results_ct}
\end{table}

\subsection{Metrics}

We evaluate the anomaly detection performance of the methods with two metrics. Firstly, we measure the area under the precision-recall curve (AUPRC) at the pixel level computed for the whole test set. AUPRC evaluates anomaly scores directly without requiring to set an operating point for each method. Secondly, we calculate $\lceil\text{Dice}\rceil$, a Dice score which measures the segmentation quality using the optimal threshold for binarization found by sweeping over possible values using the test ground truth. $\lceil\text{Dice}\rceil$ represents the upper bound for the Dice scores that would be obtainable in a more practical scenario.

\begin{figure*}[t]
  \centering
  \includegraphics[width=1.0\textwidth]{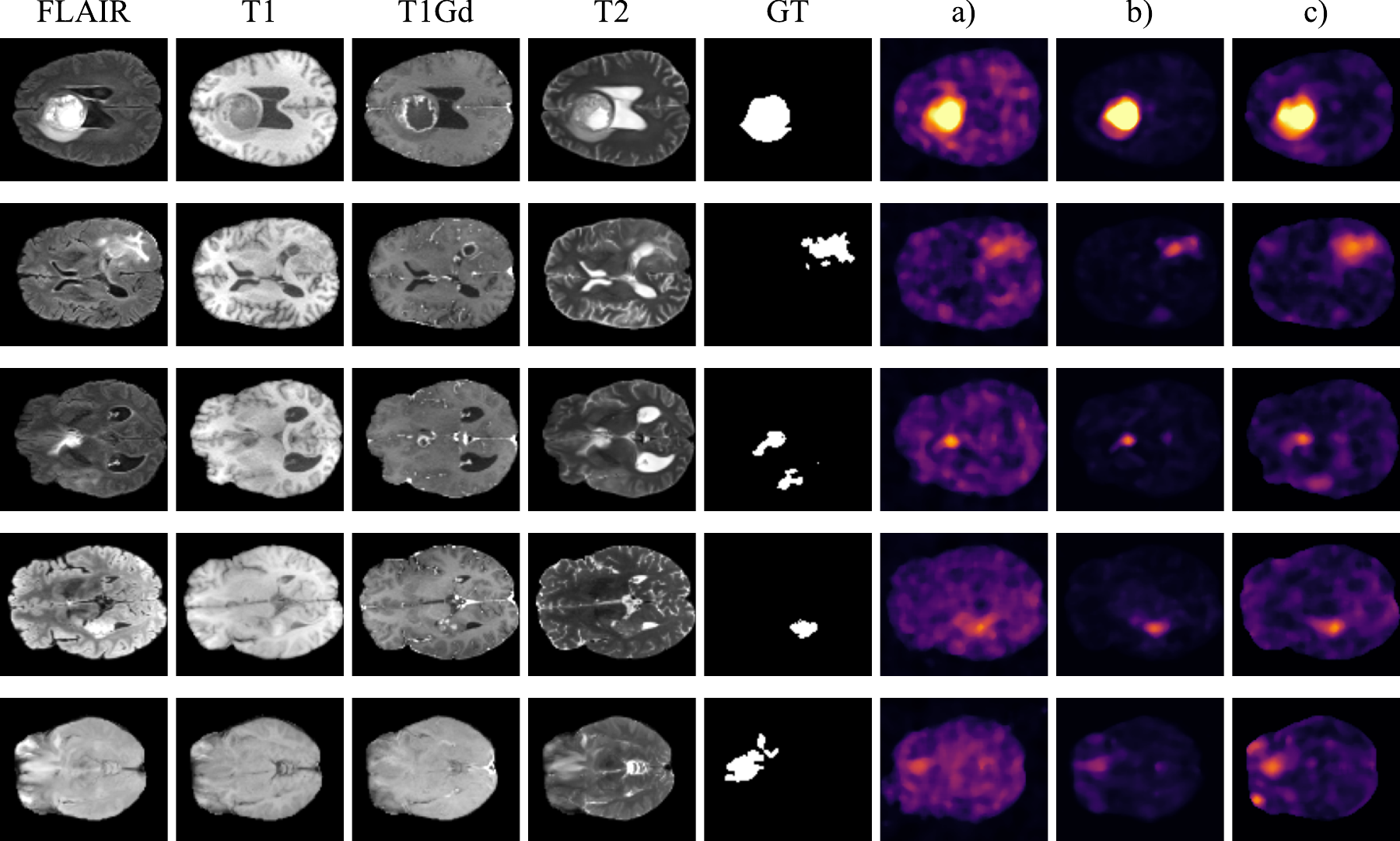}
  \caption{Sample anomaly score predictions on 2D head MRI (BraTS2021) data. Columns a), b), c) show diffusion reconstruction \citep{wyatt2022anoddpm}, KL divergence \citep{Pinaya2022} and DAE anomaly scores respectively.}
    \label{fig:qualitative_mri}
\end{figure*}

\subsection{Noise comparison}

We reported earlier (see Section \ref{sec:coarse_noise}) on experiments with the noise parameters for training a DAE, showing that the noise resolution makes a significant difference to the test-time performance of a DAE, observing similar effects across two datasets. Remarkably, the same parameters were optimal in the brain MRI and in the head CT datasets.

We now compare \emph{between} noise types, namely Gaussian noise, Simplex noise (advocated by \cite{wyatt2022anoddpm}), and our proposed coarse noise, using the optimal parameters identified in Section \ref{sec:coarse_noise} for the latter. We measure the impact on performance of the DAE and of two diffusion model inference methods proposed by \cite{Pinaya2022} and \cite{wyatt2022anoddpm}.

The results are shown in Table \ref{tab:results_diffusion_both}. Our proposed coarse noise achieves most accurate performance, significantly improving the results compared to models trained with standard Gaussian noise, and in most cases improving over models trained with Simplex noise \citep{wyatt2022anoddpm}. Interestingly for the method of \cite{Pinaya2022}, when the model is trained with Simplex or coarse noise, the intermediate KL step (similar to applying a DAE) gives better results than the subsequent diffusion inpainting step.


\subsection{Model comparison}

We compare models trained with coarse noise on the two datasets. To put the unsupervised anomaly detection performance results in context, we also provide supervised U-Net baselines, trained on a moderate number of labeled volumes.

Quantitative results can be seen in Tables \ref{tab:results_mri} and \ref{tab:results_ct}. Overall, denoising methods appear to offer more accurate anomaly detection than other unsupervised methods, with the simple DAE giving overall best performance. Diffusion models perform better than unsupervised baselines except for the simple thresholding baseline (provided for MRI but not possible for the multi-intensity lesions in CT), but worse than the DAE. As seen in Table \ref{tab:results_diffusion_both}, the intermediate KL step of \cite{Pinaya2022}'s method outperforms the results of diffusion.


\subsection{Relationship between noise resolution and anomaly size}
\label{sec:noise-vs-anomaly-size}

We further examine the relationship between the DAE noise used at training time and performance on anomalies during test time. In particular, we investigate whether the coarseness of the noise (i.e. noise resolution $\alpha$ before upsampling) has a large impact on the size of test anomalies that the DAE successfully detects as this could imply the need to tune the noise parameters for specific anomalies.

In order to isolate the effect, we evaluate DAEs trained with different noise coarseness on synthetic anomalies in 3D Head CT scans. We synthesize bright spherical anomalies inside the brain at random locations, sampling the diameter uniformly from the range of 5mm to 50mm and then multiplying the normal tissue intensity within the sphere by a factor of 2.

\begin{figure*}[t]
  
  \centering
  \includegraphics[width=1.0\linewidth]{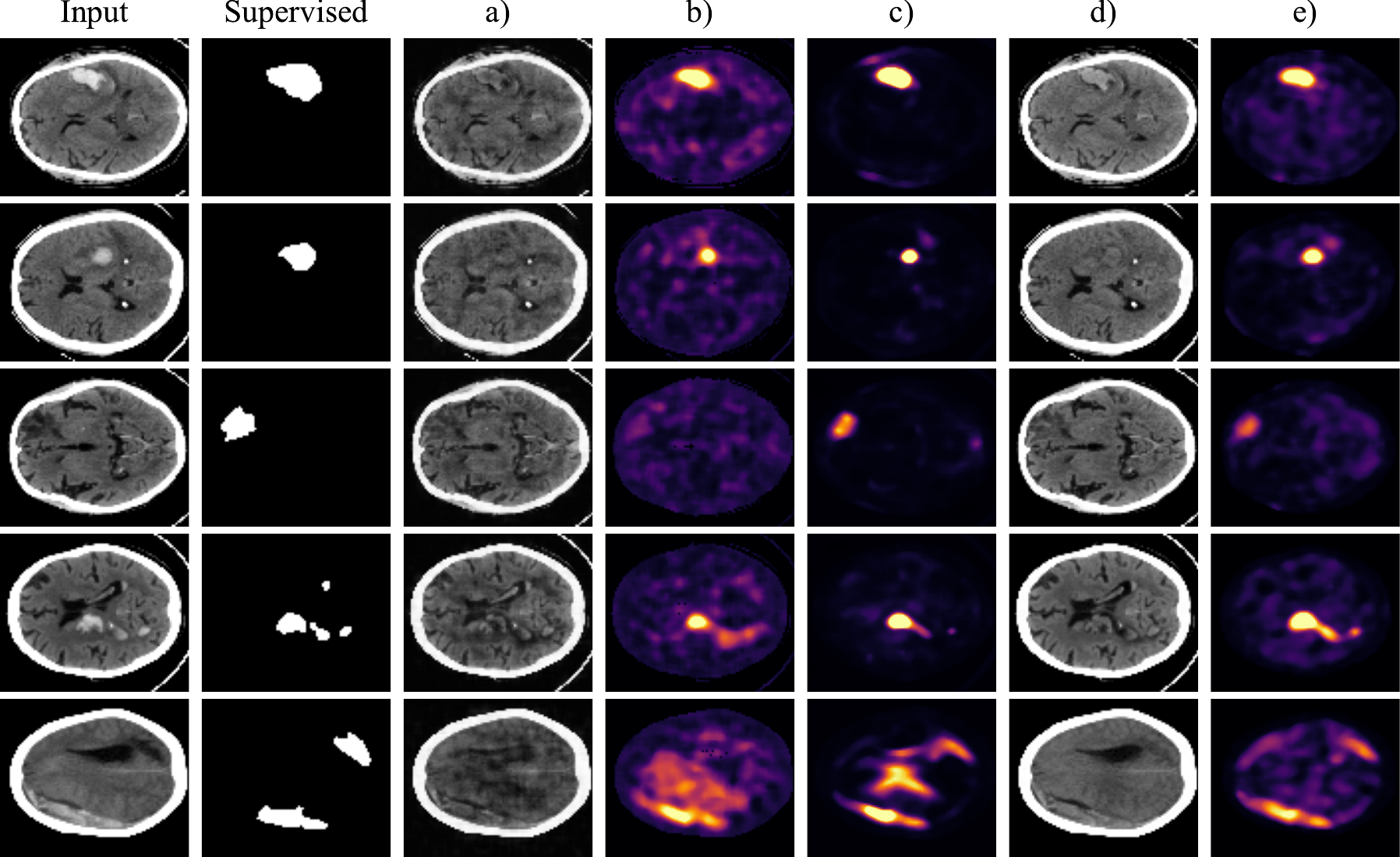}
  \caption{Sample anomaly score predictions on contrasting example slices from the 3D head CT (CQ500) data. Columns a) and b) show the show coarse noise diffusion reconstructions and anomaly scores \citep{wyatt2022anoddpm}, column c) shows the coarse noise diffusion KL divergence anomaly scores \citep{Pinaya2022}, and columns d) and e) show coarse noise DAE reconstructions and the associated anomaly scores.}
  \label{fig:qualitative_ct}
\end{figure*}

\begin{figure}[t]
  \centering
  \includegraphics[width=1.0\linewidth]{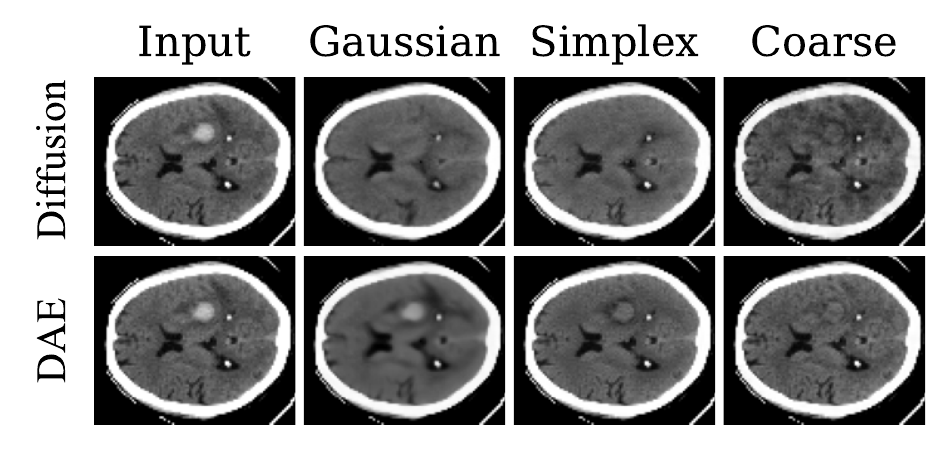}
  \caption{Comparison of reconstruction between DAE and diffusion using reconstruction procedure by \cite{wyatt2022anoddpm} across the three different noise types.}
  \label{fig:rec_samples_ct}
\end{figure}

Full results are shown in \ref{apx:relationship_noise_to_anomaly_size}. We observe no relationship between the noise resolution and the performance on anomalies in specific size range. That is, selecting a reasonably coarse noise shows consistent best performance across a wide range of synthetic anomaly sizes with no affinity towards a specific anomaly size when compared to DAEs trained with different noise coarseness parameters.

\subsection{Visualization of model outputs}

Selected examples of image slices and the corresponding denoising reconstructions and anomaly detection heatmaps for different methods are shown in Figure \ref{fig:qualitative_mri} for head MRI and Figure \ref{fig:qualitative_ct} for head CT. The DAE performs consistently well on easier tumor cases in MRI and larger hemorrhages in CT, producing strong signal predictions. Weaker signal predictions sometimes correspond to subtler tumors in MRI and ischemia cases in CT, and sometimes correspond to false positive detections.

When we examine more closely the impact of model choice and noise type on the reconstruction (see Figure \ref{fig:rec_samples_ct}), we observe the trade-off between reconstruction of the image detail vs erasure of the anomaly. The diffusion model tends to better erase the anomaly compared to the DAE, particularly when trained with Simplex or even Gaussian noise, but also yields poorer reconstruction of the image, erasing also fine details of normal anatomy such as the folds of the gyri. The DAE achieves best erasure, or at least suppression, of the anomaly when coarse noise is used, as well as best retention of the normal anatomical detail.

\section{Discussion}

\subsection{What type of noise is best?}
Our results show that the noise used for training denoising models has a large impact on anomaly detection performance. Our proposed coarse noise significantly improves the performance of both DAEs and diffusion models, outperforming naive Gaussian noise. Our coarse noise largely also outperforms the more complex Simplex noise alternative \citep{wyatt2022anoddpm}, which similarly introduces low frequencies to the noise pattern (alongside higher frequencies). Visual inspection of the reconstructions suggests that the noise used for model training impacts the trade-off between the extent of the detail that is reconstructed and the extent of erasure of anomalies, both of which contribute to better anomaly detection performance.

Remarkably, we found the same noise parameters to be optimal across the two datasets described in this paper. Further, noise parameters appeared not to be related to anomaly size, as reported in Section \ref{sec:noise-vs-anomaly-size}. Thus, it appears that noise coarseness parameters are independent of the size of test anomalies and can be set without knowledge of the nature of the target anomalies ahead of time. We note however that these datasets were processed in a similar way and are of similar anatomy (head/brain), therefore it would be desirable to do rigorous testing of many anatomies, modalities, and intensity/resolution pre-processing in order to come to a general conclusion.

\subsection{What type of model is best?}
In general, we find that employing denoising as the learning mechanism enables architectures with skip connections to be used, leading to higher fidelity reconstructions which are more effective for anomaly detection than those achieved by VAE methods with bottleneck architectures.

Considering only the denoising approaches, we find that simple denoising autoencoder (DAE) models currently outperform more advanced diffusion models across the two datasets that we evaluated, when trained with optimal noise. However, diffusion models show an exciting ability to erase anomalies and generate convincing high definition reconstructions, with the caveat that normal anatomical detail may also be erased, limiting their effectiveness at discriminating normal from abnormal. Therefore, while diffusion models are producing state-of-the-art results in image generation, further investigation is needed to find appropriate methods to apply diffusion models to anomaly detection specifically.

In terms of practical use, we note that diffusion methods (similarly to VAE restoration methods) come at a cost since all inference methods evaluated in this paper take hundreds of iterations to produce final predictions, resulting in much longer inference times than for the DAE.

\subsection{Limitations of our anomaly evaluation}
A limitation of the evaluations presented in this paper is that we have focused on a subset of anomalies which are present in the datasets, albeit also the anomalies which are of most clinical interest. This is explicit for the iCAIRD GG\&C NHS Head CT dataset, in which we annotated 3 pathologies (hemorrhage, ischemia, tumor) but extracted NLP labels for several more (see \ref{apx:patrick_labels}), filtering on these to obtain a healthy training set. 
Consequently, our metrics only approximate general anomaly detection performance. We leave comprehensive annotation of anomalies as an important avenue for future work; in fact, the performance on rarer pathologies for which expert annotations are not available is potentially more important than on common pathologies since this is where training traditional supervised approaches might be infeasible.

\subsection{Alternatives to residual error for anomaly detection}
Finally, DAEs and the diffusion inference methods we have applied rely on reconstruction error in order to detect anomalies. Reconstruction error might be suitable for prominent anomalies (e.g. large hemorrhages) but struggle with anomalies subtler in intensity contrast (e.g. ischemia). Discriminative methods (e.g. \cite{cho2021self, fpi, clfm}) that infer the anomaly score directly have been achieving success, notably in the Medical Out-of-Distribution (MOOD) Analysis MICCAI Challenge \citep{zimmerer2022mood}. They might be more suitable for subtle anomalies, since they do not use the residual error (which will be small for subtle intensity changes) as the anomaly signal; see \cite{meissen2021pitfalls} for more in-depth analysis of the pitfalls associated with using residual error. Differently to reconstruction-based methods, discriminative methods are typically trained by synthesizing abnormal data to discriminate from the healthy distribution. This has pros and cons, allowing easier application of domain knowledge about the nature of anomalies and explicit control over the definition of ``abnormal'', at the risk of losing generality and overfitting to the selected synthetic anomalies.

\section{Conclusion}

In this paper we have demonstrated the effectiveness of a simple coarse noise model in both simple classical DAEs and more complex recently proposed diffusion models for anomaly detection across two datasets. We find that the parametrization of the noise model has a wide tolerance, giving robust transfer across datasets and denoising methods. As part of this work, we implemented an anomaly detection pipeline in a real-world scenario involving the collation of a healthy training set by running NLP methods on radiology reports, thereby showing that a largely automated pipeline is possible.

Overall the classical DAE outperforms other methods, in terms of implementation simplicity, accuracy, and inference speed. While detection is successful for more obvious instances of anomalies such as tumors, hemorrhages and ischemia, further accuracy improvements are required to achieve reliable detection of subtle anomalies. Diffusion models applied to anomaly detection are as yet in their infancy and provide a promising avenue for further research.

\section*{Declaration of Competing Interest}
The authors declare that they have no known competing financial interests or personal relationships that could have influenced the work reported in this paper.

\section*{Acknowledgements}

This work is part of the Industrial Centre for AI Research in Digital Diagnostics (iCAIRD) which is funded by Innovate UK on behalf of UK Research and Innovation (UKRI) project number 104690. We thank the West of Scotland Safe Haven at NHS Greater Glasgow and Clyde for their assistance in creating this dataset. We would also like to acknowledge assistance of Canon Medical Research Europe Limited in providing the Canon Safe Haven Artificial Intelligence Platform (SHAIP) tool, assisting with the deidentification of data and the provision of a secure machine learning workspace. 

We acknowledge Engineering and Physical Sciences Research Council (EPSRC) for funding part of this work through the EPSRC Centre for Doctoral Training in Applied Photonics (CDTAP) managed by Heriot-Watt University.

This work was supported by the University of Edinburgh, the Royal Academy of Engineering and Canon Medical Research Europe via PhD studentships of Pedro Sanchez (grant RCSRF1819\textbackslash8\textbackslash25).

S.A.\ Tsaftaris acknowledges the support of Canon Medical and the Royal Academy of Engineering and the Research Chairs and Senior Research Fellowships scheme (grant RCSRF1819\textbackslash8\textbackslash25).

Many thanks to Sin Yee Foo, Harris Hameed, and Paul Donnelly from GG\&C NHS for creating the pathology annotations which we used for our evaluation.

Many thanks to Paul Thomson and Ewan Hemingway for their help with developing the imaging pre-processing pipeline.

\bibliographystyle{model2-names.bst}\biboptions{authoryear}
\bibliography{refs}

\clearpage

\section*{Supplementary Material}
\appendix

\section{DAE vs VAE reconstruction comparison}
\label{apx:rec_comparison}

\begin{figure}[h]

  \centering
  \includegraphics[width=1.0\linewidth]{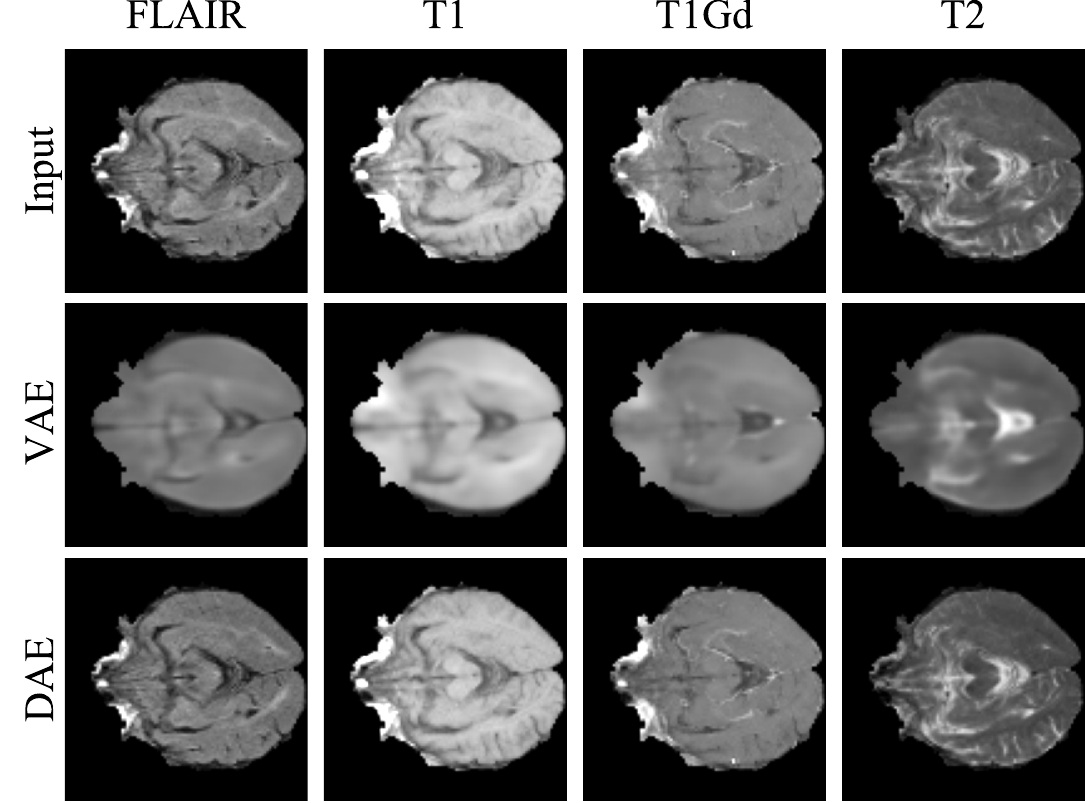}
  \caption{Sample healthy brain reconstructions from VAE and DAE models. The DAE gives more precise reconstructions. The VAE reconstruction quality could be improved by increasing bottleneck dimensionality, however this would negatively impact anomaly detection performance.}
  \label{fig:rec_comparison}
\end{figure}

\section{Neural network architectures}
\label{apx:architecture_details}

\begin{figure}[H]
  \centering
  \includegraphics[width=\textwidth]{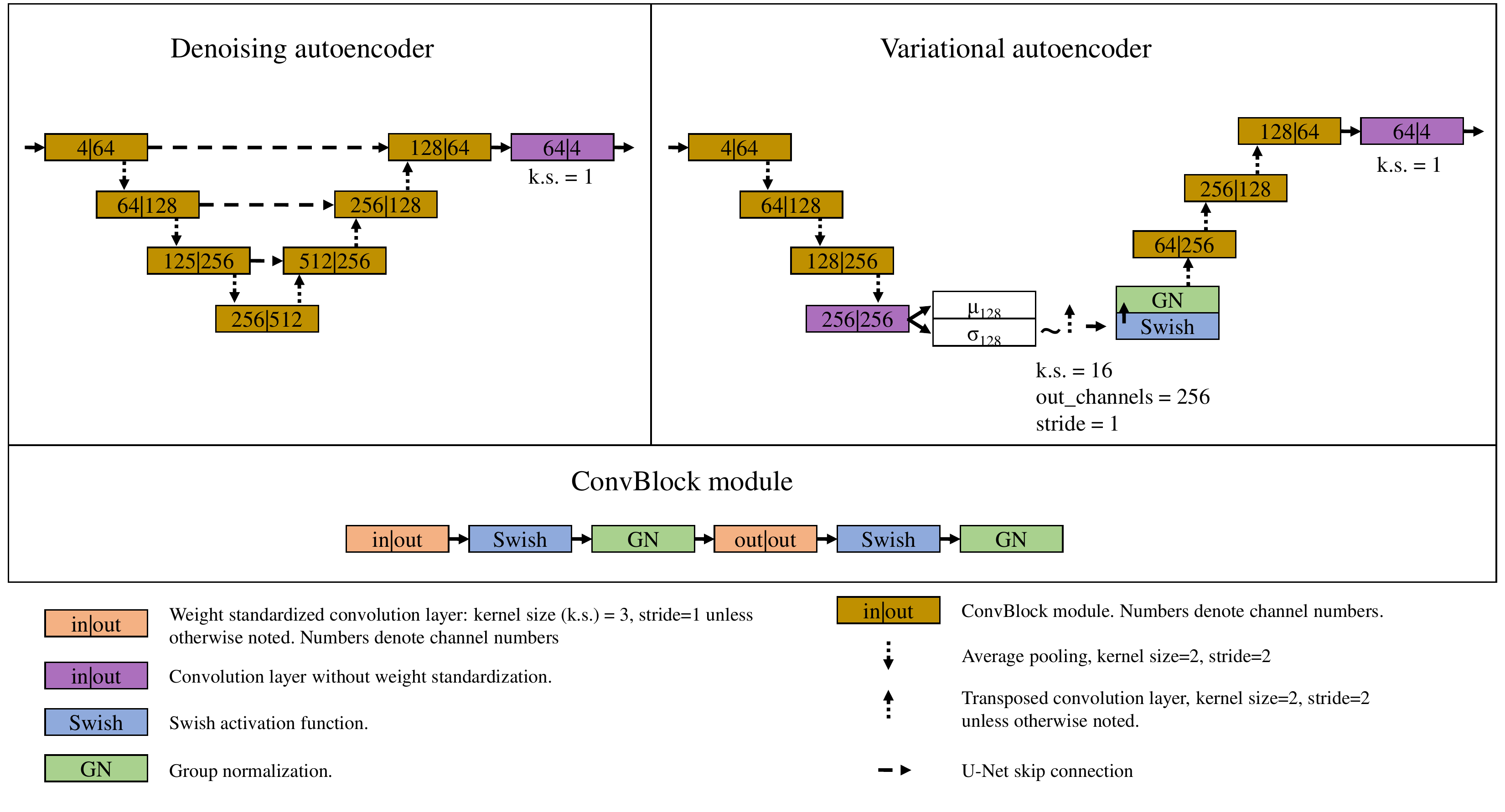}
  \caption{Architectures of 2D DAE and VAE models used in brain MRI experiments.}
  \label{fig:architecture}
\end{figure}

\clearpage

\section{Radiology report labels}
\label{apx:patrick_labels}

\begin{table}[h]
  \caption{List of report labels extracted from radiology reports using the method of \cite{patrick_templates}. We do not exclude scans with associated positive/uncertain labels which are \underline{underlined} from our healthy training set, since we decide that scans with only these labels (and no others) are ``normal for age''.}
  \label{tab:patrick_labels}
  \centering

  \begin{tabularx}{\linewidth}{X}
  \toprule

  \bfseries Radiographic findings \\
  
  artefact, collection, compression, dilation, effacement, herniation, hyperdensity, \underline{hypodensity}, loss of differentiation, malacic changes, mass effect, midline shift, oedema, swelling. \\
  
  \cmidrule(r){1-1}
  \bfseries Clinical impressions \\
  
   abscess, \underline{atrophy}, aneurysm, \underline{calcification}, cavernoma, \underline{cerebral small vessel disease}, congenital abnormality, cyst, evidence of surgery/intervention, fracture, gliosis, hemorrhage, hydrocephalus, ischemia, infection, tumor, vessel occlusion, lesion, pneumocephalus. \\

  \bottomrule
  \end{tabularx}
\end{table}

\section{Selecting and annotating the iCAIRD test set}
\label{app:icaird-test-set}

We provide additional details below on the process by which the iCAIRD test set was selected and annotated.

\vspace{5pt}\noindent{}\textbf{Scan selection:}
For hemorrhage and ischemia cases, our primary source was the Scottish Stroke Care Audit (SSCA) records for which we had access for the stroke episodes in the dataset; we searched these records for stroke episodes classed as ``hemorrhagic'', ``ischemic'', or ``hemorrhagic transformation''. For cases of tumors and rarer hemorrhages (epidural and subdural), we used a combination of ICD-10 code and free text searches of the Scottish Morbidity Records (SMRs) and radiology reports (e.g., ``extradural'', ``extra-dural'', ``extra dural'', ``epidural'', ``edh'', ``subdural'', and ``sdh''), respectively. We then excluded
scans acquired prior to 2016 for image compression reasons.

\vspace{5pt}\noindent{}\textbf{Annotation and Review:} We recruited 3 GG\&C clinicians (one Consultant Neuroradiologist and two senior Radiology trainees) to perform pixel-level annotation, following the annotation protocol prepared for this project. For the selected cases, all hemorrhage, ischemia and tumor lesions present were annotated, including any surrounding regions of edema for hemorrhagic lesions.
The Consultant Neuroradiologist acted also as reviewer;
40\% of cases were randomly selected for review
and annotators also had the option of sending any of the remaining 60\% for review when they required a second opinion.


\newpage
\section{Relationship between noise coarseness and anomaly size}
\label{apx:relationship_noise_to_anomaly_size}

\begin{figure}[h]
  \centering
  \includegraphics[width=0.9\linewidth]{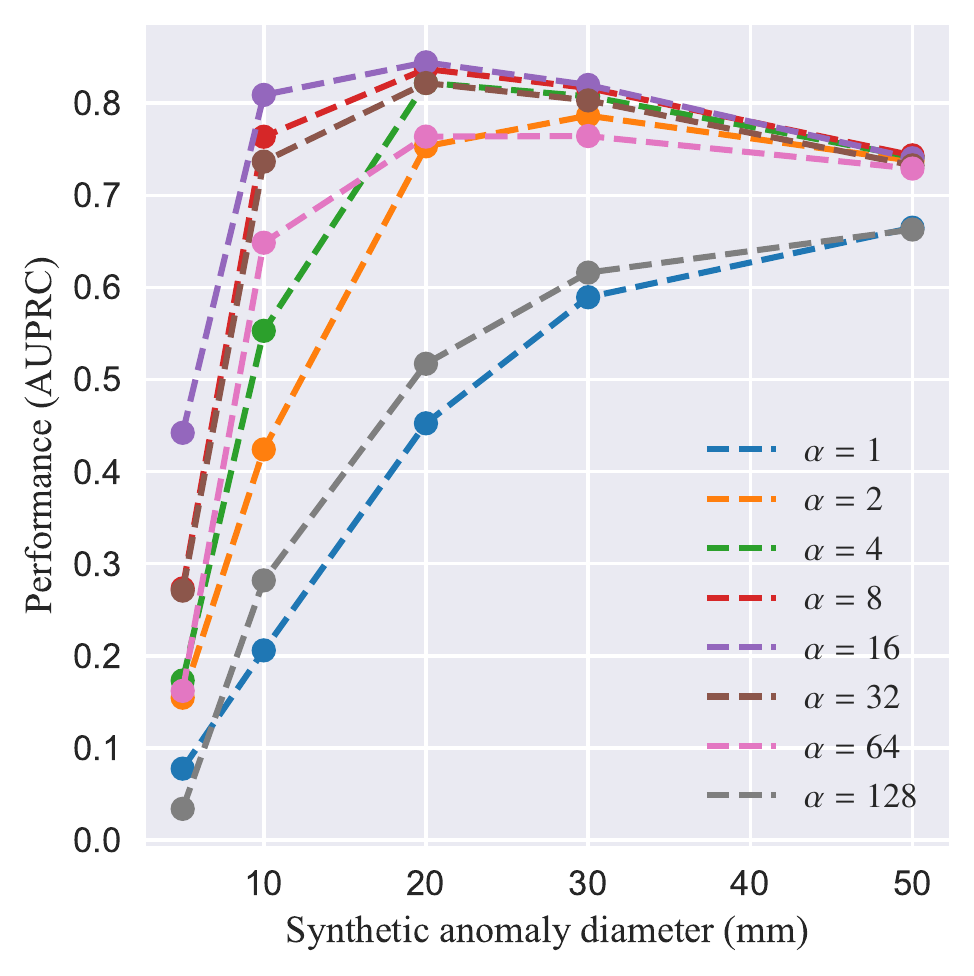}
  \caption{DAE anomaly detection performance evaluated using synthetic anomalies of different sizes with models trained with noise generated at different resolutions $\alpha$. Each point indicates a mean of three runs. The best noise resolution ($\alpha=16$) generalizes the best across a wide variety of synthetic anomaly sizes.}
  \label{fig:anomaly-size}
\end{figure}

\end{document}